\documentclass[reprint, twocolumn, superscriptaddress, notitlepage]{revtex4-1}
\usepackage{amsfonts}
\usepackage{amssymb}
\usepackage[english]{babel}
\usepackage{hyperref}   % use for hypertext links, including those to external documents and UrLs
\usepackage{color}
\usepackage{float}
\usepackage{placeins}
\usepackage{amsmath}
\usepackage{hyphenat}
\usepackage{graphicx}  
\usepackage{caption}
\usepackage{subcaption}
\usepackage{siunitx}
\usepackage[symbol]{footmisc}

\usepackage{natbib}
\newcommand{\bn}{\mbox{\boldmath $n$}}
\newcommand{\bN}{\mbox{\boldmath $N$}}

\newcommand{\bx}{\mbox{\boldmath $x$}}
\newcommand{\bu}{\mbox{\boldmath $u$}}

\newcommand{\bB}{\mbox{\boldmath $B$}}
\newcommand{\br}{\mbox{\boldmath $r$}}

\newcommand{\bj}{\mbox{\boldmath $j$}}
\newcommand{\be}{\mbox{\boldmath $e$}}
\newcommand{\bv}{\mbox{\boldmath $v$}}

\newcommand{\bG}{\mbox{\boldmath $G$}}
\newcommand{\bD}{\mbox{\boldmath $D$}}

\begin{document}
\raggedbottom %ensure even spacing of paragraphs
\title{Magnetic skyrmion binning}

\author{Charles Kind}
%\email{charles.kind@bristol.ac.uk}
\affiliation{School of Mathematics, University of Bristol, Bristol BS8 1TW, UK}
\author{David Foster}
\affiliation{UKAEA, Culham Centre for Fusion Energy, Abingdon, OX14 3DB, UK}
%\affiliation{ }

\date{\today}

\begin{abstract}
When spin polarised electrons flow through a magnetic texture a transfer torque is generated. We examine the effect of this torque on skyrmions and skyrmion bags, skyrmionic structures of arbitrary integer topological degree, in thin ferromagnetic films. Using micromagnetic simulations and analysis from the well known Thiele equation we explore the potential for sorting or binning skyrmions of varying degrees mechanically. We investigate the applicability of the Thiele equation to problems of this nature and derive a theory of skyrmion deflection ordered by topological degree. We show that skyrmions and skyrmion bags have the potential to move in different directions under constant current which has significant potential for technical applications.
\end{abstract}

\maketitle
\section{Introduction}
Miniaturisation of magnetic storage devices has reached two fundamental limits. On one hand the reduced lateral dimensions lowers the stability of magnetic domains. On the other hand many devices still suffer comparatively large power consumption. This is economically less favourable, ecologically undesirable and further diminishes the stability of nano scale devices. A possible route to alleviate some of these issues is the utilisation of magnetic skyrmions, exotic spin textures with some level of topological protection.

The occurrence of skyrmions was first predicted at the end of the last century
\cite{BogdanovThermodynamicallystablevortices1989, BogdanovThermodynamicallystablemagnetic1994, RosslerSpontaneousskyrmionground2006}
, and they were later experimentally observed by neutron scattering \cite{MuhlbauerSkyrmionLatticeChiral2009}, by Lorentz transmission electron microscopy \cite{YuRealspaceobservationtwodimensional2010} and by spin-resolved scanning tunnelling microscopy \cite{HeinzeSpontaneousatomicscalemagnetic2011}.

It has been shown that such configurations are more stable than conventional ferromagnetically ordered systems but most importantly they can be driven with low electric currents reducing the power consumption. Various basic technologies have been proposed, \cite{FertSkyrmionstrack2013, ZhangMagneticskyrmionlogic2015} to use magnetic skyrmions to encode data by skyrmion separation. But this has a fundamental problem, namely that experimentally produced materials contain defects and inhomogeneous regions which can effect skyrmion motion, causing data to become lost. An elegant solution to this could be the skyrmion bag.

Skyrmion bags, or sacks, \cite{FosterTwodimensionalskyrmionbags2019, RybakovChiralmagneticskyrmions2019} are nested skyrmionic structures of any integer topological degree. The textures are composed of a single skyrmion outer boundary and then a number of inner antiskyrmions,
which can themselves contain skyrmions. In the two dimensional (2D) model describing thin films, bags can be characterised by their total topological degree defined as

\begin{align} \label{TopDeg}
Q = \frac{1}{4 \pi} \int \bn \cdot \left( \partial_1 \bn \times \partial_2 \bn \right) \: d^2 x,
\end{align}

where $\bn \left( \bx \right)$ is the unit vector field of magnetization. Under this definition, a single skyrmion has the degree $Q=-1$ and an antiskyrmion $Q=1$.

The notation we use to describe the simple skyrmion bags, possessing only a number of individual antiskyrmions inside the outer skyrmion, is $S \left( N \right)$, where $n$ is the number of antiskyrmions and the total degree is therefore $Q = N - 1$ \cite{FosterTwodimensionalskyrmionbags2019, KindExistencestabilityskyrmion2020}. We use $S(0)$ as the label for a single skyrmion.

Encoding data using skyrmion bags in a current driven racetrack device has been proposed \cite{FosterTwodimensionalskyrmionbags2019} and here we consider the effects of spin polarised current on those bags. When spin-polarised current is introduced to a ferromagnetic system spin transfer torques (STT) are generated which affect the magnetisation. As skyrmions are propelled by electric current, they can become deflected in a direction orthogonal to the current, an effect known as the skyrmion hall effect. Notably, skyrmion bags have the potential to deflect in opposing directions to skyrmions.

%\onecolumngrid

\begin{figure}[h!]
\centering
\includegraphics[width = 0.90\columnwidth]{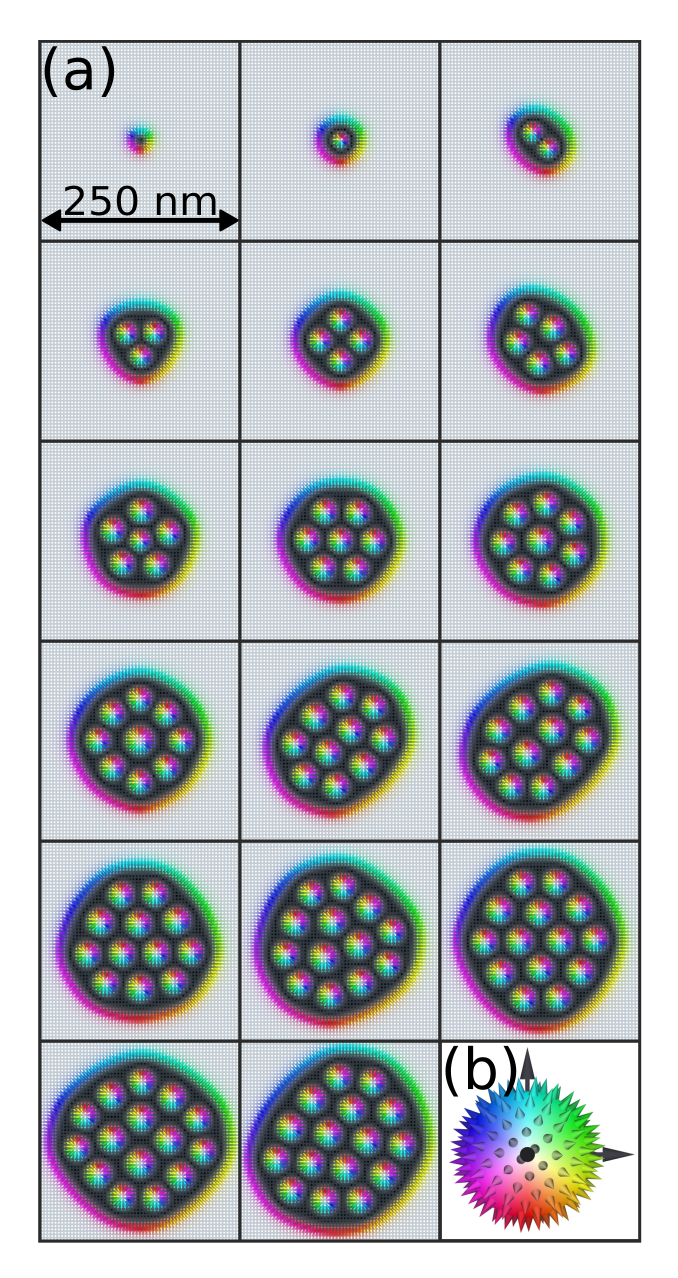}
\caption{\textbf{A skyrmion and skyrmion bags.}
   \textbf{(a)} Mumax3 simulated $S(0)$ to $S(16)$ bags in $250$ nm$^2$ boxes for constants $M_{s} = 580$ kAm$^{-1}$, exchange $15$ pJm$^{-1}$, interfacial DMI $3.4$ mJm$^{-2}$ and uniaxial anisotropy $0.8$ MJm$^{-3}$ in the $+z$ direction.
   \textbf{(b)} The Runge colour sphere used by our software to represent directions on $S^2$.
  }
\label{fig1}
\end{figure}

%\twocolumngrid

\section{Theory}

We adopt the Thiele approach for the collective coordinates of magnetisation textures. We suppose the stationary limit, where the magnetisation texture moves with constant velocity, and assume that the texture does not deform. We consider both current-in-plane (CIP) and current perpendicular-to-plane (CPP) geometries using the Zhang-Li \cite{ZhangRolesNonequilibriumConduction2004b} and Slonczewski \cite{SlonczewskiCurrentdrivenexcitationmagnetic1996} models respectively. The travelling wave ansatz, $\bn \left( \br , t \right) = \bn \left( \br - \bv_d t\right)$ , is then applied to the Landau-Lifshitz-Gilbert \cite{Gilbertphenomenologicaltheorydamping2004} dynamical equation with appropriate torque to produce \cite{ThiavilleMicromagneticunderstandingcurrentdriven2005a, EverschorCurrentinducedrotationaltorques2011, EliasSteadymotionskyrmions2017}
Thiele's equation \cite{ThieleSteadyStateMotionMagnetic1973}. In the CIP geometry we have

\begin{align} \label{ThieleCIP}
\bG \times  \left( \bj - \bv \right) + \bD \left( \beta \bj - \alpha \bv \right) = 0,
\end{align}

where $\bG = 4 \pi Q \be_z$ is the gyrocoupling vector, 
%$Q$ is the topological degree from Eq. (\ref{TopDeg}), 
$\bD$ is the dissipative tensor with components 

\begin{align} \label{DTensor}
D_{ij} = \int \partial_i \bn \cdot \partial_j \bn \: d^2x, 
\end{align}

$\bv$ is the velocity of the centre of mass of the skyrmion, $\bj$ is the current vector, $\alpha$ is the Gilbert damping constant and $\beta$ is the dissipative spin transfer torque parameter. For an axially symmetric texture the dissipative tensor $\bD$ will have components $D_{1,1} = D_{2,2}$ and $D_{i,j} = 0$, for $i \ne j$  \cite{HuberDynamicsspinvortices1982}. 

In the CPP geometry the Thiele equation is  \cite{ZhangControlmanipulationmagnetic2016},

\begin{align}\label{ThieleCPP}
\bG \times \bv - \alpha \bD \cdot \bv + \mathcal{B} \cdot \bj = 0,
\end{align}

where $\mathcal{B}$ is the driving force term linked to the spin transfer torque with the tensor 

\begin{align} \label{BTensor}
\mathcal{B} &= \frac{u}{aj} \begin{pmatrix} -\mathcal{I}_{xy} & \mathcal{I}_{xx} \\ -\mathcal{I}_{yy} & \mathcal{I}_{yx} \end{pmatrix},
\end{align}

where the components of Eq. (\ref{BTensor}) are

\begin{align} \label{BComp}
\mathcal{I}_{ij} &= \int \left( \partial_i \bn \times \bn \right)_j dx \: dy,
\end{align}

$u = |\frac{\gamma_0 \hbar}{\mu_0 e}|\frac{jP}{2 M_s}$ is the STT coefficient where $P$ is the spin polarisation rate, $\hbar$ is the reduced Planck constant, $e$ is the electron charge, $j$ is the applied current density, $\gamma_0$ is the gyromagnetic ratio, $\mu_0$ is the vacuum permeability constant and $M_s$ is the saturation magnetisation. Notably, for an axially symmetric solution $\mathcal{I}_{yy} = \mathcal{I}_{xx} = 0$, $\mathcal{I}_{xy} = -\mathcal{I}_{yx}$ \cite{ZhangControlmanipulationmagnetic2016}.
All other terms are the same as in Eq. (\ref{ThieleCIP}) We ignore pinning effects in this analysis.

The gyrocoupling vector, $\bG$, in Eq.(\ref{ThieleCIP}) and Eq.(\ref{ThieleCPP}) models an effective Magnus force from the electrons picking up a geometric Berry phase and pushing the skyrmion perpendicular to it's direction of motion resulting in a topological Hall effect \cite{NeubauerTopologicalHallEffect2009}. This effect is proportional to the topological degree $Q$. We theorise that skyrmion bags \cite{FosterTwodimensionalskyrmionbags2019}, with different topological numbers $Q$, could be separated simply by the effect their topological degree has on their trajectory.

Using a Bogomlnyi type argument \cite{MantonTopologicalSolitons2004}, we find a lower bound on the dissipation tensor by integrating the inequality
%We find a lower bound on the dissipation tensor by considering a Bogomlnyi type argument from the $\mathbb{CP}^1$ lump model \cite{MantonTopologicalSolitons2004} where $\bn : \mathcal{S}^2 \mapsto \mathcal{S}^2$. 

%The homotopy group is $\pi_2 \left( \mathcal{S}^2 \right) = \mathbb{Z}$ implying that each field is characterized by a topological charge $Q \in \mathbb{Z}$ represented by Eq.(\ref{TopDeg}) which is the pull back of the area form on $\mathcal{S}^2$. With a static field energy of

%\begin{align} \label{BogEnergy}
%D = \frac{1}{2} \int \partial_i \bn \cdot \partial_i \bn \: d^2 x,
%\end{align}

%which is the dissipation tensor $\bD$ from Eq. (\ref{Thiele}) in the form $D \mathbb{I}$, where $\mathbb{I}$ is the identity matrix in $\mathbb{R}^2$, and by integrating the inequality

\begin{align} \label{BogIneq}
\left( \partial_i \bn \pm \epsilon_{ij} \bn \times \partial_j \bn \right) \cdot
\left( \partial_i \bn \pm \epsilon_{ik} \bn \times \partial_k \bn \right) > 0,
\end{align}

over the plane and using Eq's. (\ref{TopDeg}) and (\ref{DTensor}). This gives a  bound on $D$

\begin{align} \label{BogBound}
D \ge 4 \pi |Q|,
\end{align}

where we have assumed that the skyrmion bags are approximately axially symmetric  hence $D_{1,1} = D_{2,2}$ and $D_{i,j} = 0$ for all $Q$. %(see Eq. (\ref{TopDeg}))%\cite{HuberDynamicsspinvortices1982}

The diagonal dissipation tensor \cite{HuberDynamicsspinvortices1982} integrand is everywhere greater or equal to zero due to it's inner product. By assuming that around every skyrmion or antiskyrmion there can be drawn a line through the plane where the magnetisation is the same as the far field, $+\be_z$, we separate the plain into compactly supported textures and assume each approximates a free (anti)skyrmion. We can therefore split the skyrmion bag into an outer skyrmion with $Q=-1$, and $N$ inner antiskyrmions. We assume that the inner antiskyrmions are all similar. Hence for a total dissipation tensor of $D = \int \partial_1 \bn \cdot \partial_1 \bn \: d^2 x$ we have

\begin{align}
D &= \int_{outer} \partial_1 \bn \cdot \partial_1 \bn \: d^2 x + N \int_{inner} \partial_1 \bn \cdot \partial_1 \bn \: d^2 x, \\
&= K + NS, \label{KNS}
\end{align}

where $K$ is the outer bag contribution and $NS$ the inner number of antiskyrmions multiplied by their, equal, diagonal dissipative terms.

If we now solve Thiele's equation in the CIP geometry (\ref{ThieleCIP}) setting $\bv_s = \left( \bj_x, 0 \right)^T$, to simulate a current moving in the $\be_x$ direction exclusively, we get the texture motion to be,

\begin{align}
\bv \cdot \be_x &= \frac{\left( \frac{D}{G} \right)^2 \alpha \beta + 1}{\alpha^2 \left( \frac{D}{G} \right)^2 + 1} \bj_x, \label{VdxCIP} \\
\bv \cdot \be_y &= \frac{\left( \alpha - \beta \right) \frac{D}{G}}{\alpha^2 \left( \frac{D}{G} \right)^2 + 1} \bj_x.  \label{VdyCIP}
\end{align}

%Here $G=4 \pi Q$ and $Q=N-1$ for the simple skyrmion bag. 
Clearly for $\alpha = \beta$ the perpendicular motion is zero. The an angle of deflection is therefore

\begin{align}
\theta_{CIP} = \mbox{arctan} \left( \frac{\left( \frac{D}{G} \right)^2 \alpha \beta + 1}{\left( \alpha - \beta \right) \frac{D}{G}} \right),
\end{align}

Using our assumption Eq. (\ref{KNS}) we see that

\begin{align} \label{DoverG}
\frac{D}{G} = \frac{K + NS}{4 \pi \left( N - 1 \right)},
\end{align}

which, in the limit of large N, goes to

\begin{align}
\lim_{N \to \infty} \frac{D}{G} = S/4 \pi.
\end{align}

For the CPP case we solve in a similar fashion and obtain,

\begin{align}
\bv \cdot \be_x = \frac{u}{aj} \frac{\alpha D \mathcal{I}_{xy}}{\left( \alpha D \right)^2 + G^2}, \bj_x \label{VdxCPP} \\
\bv \cdot \be_y = \frac{u}{aj} \frac{G \mathcal{I}_{xy}}{\left( \alpha D \right)^2 + G^2} \bj_x. \label{VdyCPP}
\end{align}

%In this case we consider the ratio of velocities, as our primary concern in this study is t
The angle of texture deflection is,

\begin{align}
\theta_{CPP} = \mbox{arctan} \left( \frac{\alpha D}{G} \right),
\end{align}

and in the limit we have

\begin{align}
\lim_{N \to \infty} \frac{\alpha D}{G} = \alpha S/4 \pi.
\end{align}

We therefore see that the contribution to perpendicular motion will, in both the CIP and CPP geometries, reach a limit. $S$ itself has a positive lower bound from Eq. (\ref{BogBound}) and here $|Q|=1$ as an individual antiskyrmion has topological charge $-1$. Hence we expect the differences in perpendicular motion between simple bags of higher degree to decrease towards a limit, making binning of bags with higher topological degrees theoretically impractical for a material with these properties. 
%larger numbers of inner antiskyrmions levels theoretically impractical.

%\begin{figure}[H]
%\centering
%\includegraphics[width = 1\columnwidth]{figs/doverqline.jpeg}
%\caption{\textbf{Plot of $D/Q$}
%  }
%\label{fig1}
%\end{figure}

\section{Simulations}
To investigate this we simulated the motion of simple skyrmion bags, of degrees $N = 0,1,2,...,16$, in a large material under the influence of a current. 

The simulations where performed using the GPU-accelerated micromagnetic simulation program MuMax3 \cite{VansteenkistedesignverificationMuMax32014} with Landau-Lifshitz dynamics in the form
\begin{align}\label{llg}
\frac{\partial \bn}{\partial t}=\gamma\frac{1}{1+\alpha^2}\left(\bn \times \bB_{\mathrm{eff}}+\alpha\bn
\times\left( \bn \times \bB_{\mathrm{eff}} \right) \right),
\end{align}
where $\gamma \approx 176$ rad(ns$\:$T)$^{-1}$ is the electron gyromagnetic ratio, $\alpha=0.3$ the dimensionless damping parameter, $\bB_{\mathrm{eff}} = \delta E / \delta \bn$ the effective field and $\bn(\bx)=\bN (\bx)/M_{s}$ the magnetisation vector field normalised by the saturation magnetisation. The initial bag configurations were built from template functions of individual skyrmions and antiskyrmions, in the $S(0)$ to $S(16)$ configurations. 
%Here the notation $S(0)$ defines to a single skyrmion. 
%The simulations showed a significant drop in energy from the initial condition which indicates the system achieving stability. 
These bags where chosen as sixteen distinct configurations allow for the encoding of 4 bits of data.

The simulation geometry is a $1024\times1024$ nm$^2$ square of $1$ nm thickness, in order to represent a typical wire that could be fabricated using lithographic processing. Cell size of $2 \times 2 \times 1$ nm$^3$ has been used. Material parameters are: saturation magnetisation $M_{s} = 580$ kAm$^{-1}$, exchange $15$ pJm$^{-1}$, interfacial DMI $3.4$ mJm$^{-2}$ and uniaxial anisotropy along the $+z$ direction $0.8$ MJm$^{-3}$. These approximate a Pt/Co system with high interfacial DMI \cite{MetaxasCreepFlowRegimes2007}. 

In the CIP geometry the non-adiabacity of spin transfer torque constant is set to $\xi = 0.2$ and the Zhang-Li \cite{ZhangRolesNonequilibriumConduction2004b} model is used by MuMax3 to simulate the torque from in plane polarised current flow in the $\be_x$ direction: 

\begin{align}\label{ZL}
\overrightarrow{\tau}_{ZL} &=\frac{1}{1+\alpha^2} \left( \left( 1 + \xi \alpha \right)\bn \times \left( \bn \times \left( \bu \cdot \nabla \right) \right) \bn  \nonumber \right. \\
&+ \left. \left( \xi - \alpha \right) \bn \times \left( \bu \cdot \nabla \right) \bn \right) \\
\bu &= \frac{\mu_B \mu_0}{2 e \gamma_0 B_{sat} \left( 1 + \xi^2 \right)}\bj
\end{align}

where $\bj \cdot \be_x = 50$ nA$\:$nm$^{-2}$  is the current density, $\mu_B$ the Bohr magneton and $B_{sat}$ the saturation magnetisation expressed in Tesla.

In the CPP geometry we have:

\begin{align}\label{Slonc}
\overrightarrow{\tau}_{SL} &= \beta \frac{\epsilon - \alpha \epsilon '}{1 + \alpha^2} \left( \bn \times \left( \bn_p \times \bn \right) \right) \nonumber \\
&- \beta \frac{\epsilon ' - \alpha \epsilon}{1 + \alpha^2} \bn \times \bn_p \\
\beta &= \frac{j_z \hbar}{M_s e d} \\
\epsilon &= \frac{P \left( \overrightarrow{\br} , t \right) \Lambda^2}{\left( \Lambda^2 + 1 \right) + \left( \Lambda^2 - 1 \right)  \left( \bn \cdot \bn_p \right)} 
\end{align}

where $j_z = 10$ nA$\:$nm$^{-2}$ is the current density along the $z$-axis, $d$ the free layer thickness, $\bn_p = -\be_y$ the fixed layer magnetisation, $P=0.4$ the spin polarisation with $\Lambda = 1$ the Slonczewski parameter and $\epsilon ' = 0$ the secondary spin torque parameter.

The simulations used Mumax3's periodic boundary conditions and were run for $1$ $\mu$s with the vector field recorded every $10$ ns to obtain the centre of mass of the skyrmionic texture and to calculate the tensors $\bD$ and $\mathcal{B}$ given by Eq. (\ref{DTensor}) and the total positive and negative contributions to the topological degree (Eq. (\ref{TopDeg})).

%To calculate the Thiele trajectories we developed a numerical integrator to evolve the Thiele equations, a unique feature of our integrator is that
For the Thiele trajectories we not only changed the skyrmions topological degree, we also changed the dissipative tensor, $D$, as a function of $Q$.
%constant $ D = \frac{1}{2} \int \pa_i \bn \cdot \dot \pa_i \bn d^2x$. 
This is an important feature because as the skyrmion bag grows so does $D$, which leads to a reduced deflection. This is why,  bags which contain a large number of skyrmions, have similar trajectories. 

\section{Results}

\begin{figure}[h!]
\centering
\includegraphics[width = 1\columnwidth]{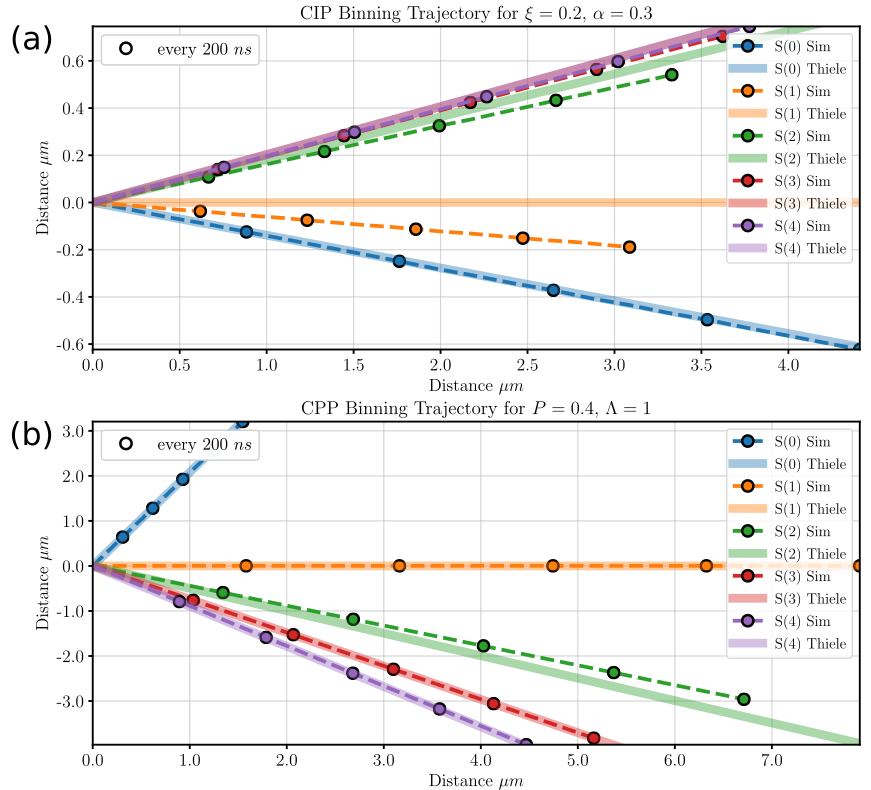}
\caption{\textbf{Binning trajectories in simulation and theory.}
	Here we only plot up to the $S(4)$ bag as the higher degree skyrmion bags converge at around the $S(4)$ trajectory.
   \textbf{(a)} CIP binning trajectories for $S(0)$ to $S(4)$.
   \textbf{(b)} CPP binning trajectories for $S(0)$ to $S(4)$.
  }
\label{fig2}
\end{figure}

The centre of mass of each texture is computed at every time step and we plot their trajectories in Fig. (\ref{fig2}) alongside the trajectories calculated by our Thiele equations.

\onecolumngrid

\twocolumngrid

We find that the Thiele equation accurately predicts the path of the skyrmion and at higher topological degrees the limiting $e_y$ displacement of the skyrmion bags. The bags with the largest off diagonal $\bD$ tensor components deviated from the Thiele predicted paths the most and where also the least radially symmetric. The $S(1)$ bag in the CIP geometry deviated significantly from the Thiele prediction of zero deflection from the x-axis with an angle of $-3.49$ degrees, see table (\ref{tab:deflections}).

\begin{figure}[h!]
\centering
\includegraphics[width = 0.95\columnwidth]{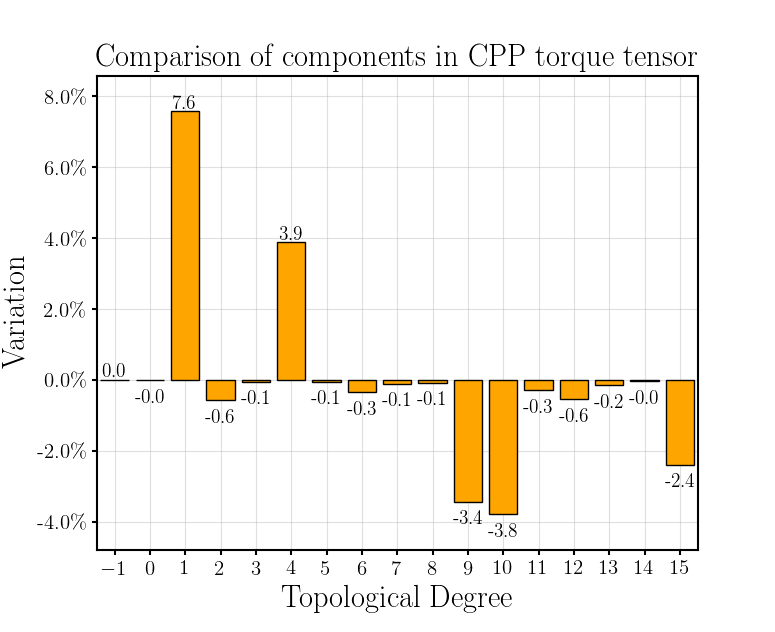}
\caption{\textbf{Comparison of $\bD$ tensor components in the CPP geometry.} The magnitude of the average of the off diagonal tensor components $\left(D_{1,2} + D_{2,1}\right)/2$ vs the diagonal components $\left(D_{1,1} + D_{2,2}\right)/2$ expressed as a percentage for	 $S(0)$ to $S(16)$. The tensor components, eq (\ref{DTensor}), were numerically integrated from MuMax3 simulation data. The CPP case is plotted as the CIP data varies by less than $1$ \%.
  }
\label{fig3}
\end{figure}

We numerically compute the $D_{i,j}$ components for the dissipative tensor from Eq. (\ref{DTensor}) and plot the ratio of the diagonal to off diagonal components, see Fig. (\ref{fig3}). We find that the bags with the least radial symmetry have the largest off diagonal components with the $S(2)$ bag being the least symmetric by this measure; compare Fig. (\ref{fig3}) with Fig. (\ref{fig1}). The off diagonal components of the $S(2)$ bag are less than $1/10$th the size of the diagonal components with all other bags having a ratio of at most $1/25$. These values vary by less than one part in $10000$ throughout the simulations indicating that the textures do not appreciably deform under the applied currents. It is this `approximate' symmetry which allows us to derive the bounds in Eq. (\ref{BogBound}).

We numerically integrate the topological degree, Eq. (\ref{TopDeg}), and find that the outer skyrmions topological degree increases as $N$ increases. Also the contribution from the inner antiskyrmions grow as $N$ increases. The inner antiskyrmions increase by around $0.1\%$ from $S(1)$ to $S(2)$ decreasing to $0.003\%$ from the $S(15)$ to $S(16)$ bag whilst the outer skyrmions degree increases close to constantly by $0.15\%$ per addition of an antiskyrmion to the bag.

In the CIP geometry the fastest skyrmion is the single skyrmion with the speed of bags increasing as their degree increases. In CPP, however, the single skyrmion is slowest with the $S(1)$ skyrmionium being the fastest \cite{GobelElectricalwritingdeleting2019} and the bags speeds decreasing as their topological degree increases
%with the magnitude of the bags velocity decreasing by degree but always being faster then the single $S(0)$ skyrmion
, see table (\ref{tab:deflections}).\\

\begin{table}[H]
  \begin{center}
    \caption{Texture deflection angle and speed for each bag by geometry (2 d.p.).}
    \label{tab:deflections}
    \begin{tabular}{c|c|c|c|c|c|c|c}
      \textbf{ Bag } & \textbf{ Q } & \textbf{ CIP } & \textbf{ CIP } & 			\textbf{ CPP } & \textbf{ CPP } & \textbf{ CIP } & \textbf{ CPP }\\
      & (\ref{TopDeg}) & Sim & Thiele & Sim & Thiele & Sim & Sim\\
      & & Deg. & Deg. & Deg. & Deg. & ms$^{-1}$ & ms$^{-1}$\\
      \hline \vspace{-3mm}
      & & &\\ 
      $S(0)$ &-1& -8.04 & -7.9 & 64.23 & 64.34 & 4.41 & 3.56\\
      $S(1)$ &0& -3.49 & 0 & 0.01 & 0 & -3.09 & 7.91 \\
      $S(2)$ &1& 9.22 & 10.29 & -23.8 & -26.5 & 3.33 & 7.34\\
      $S(3)$ &2& 10.94 & 11.48 & -36.53 & -36.53 & 3.63 & 6.43\\
      $S(4)$ &3& 11.19 & 11.49 & -41.62 & -41.79 & 3.78 & 5.98\\
      $S(5)$ &4& 11.21 & 11.3 & -42.95 & -45.09 & 3.84 & 5.77\\
      $S(6)$ &5& 10.96 & 11.09 & -47.18 & -47.37 & 3.94 & 5.43\\
      $S(7)$ &6& 10.83 & 10.9 & -48.8 & -49.09 & 3.99 & 5.26\\
      $S(8)$ &7& 10.68 & 10.73 & -50.23 & -50.35 & 4.03 & 5.11\\
      $S(9)$ &8& 10.54 & 10.57 & -51.28 & -51.43 & 4.06 & 4.97\\
      $S(10)$ &9& 10.38 & 10.46 & -54.13 & -52.21 & 4.1 & 4.82\\
      $S(11)$ &10& 10.26 & 10.33 & -55.06 & -52.98 & 4.13 & 4.71\\
      $S(12)$ &11& 10.23 & 10.22 & -53.46 & -53.62 & 4.12 & 4.75\\
      $S(13)$ &12& 10.15 & 10.14 & -54.23 & -54.09 & 4.14 & 4.65\\
      $S(14)$ &13& 10.25 & 10.14 & -54.9 & -54.11 & 4.15 & 4.71\\
      $S(15)$ &14& 9.89 & 9.92 & -54.46 & -55.3 & 4.16 & 4.5\\
      $S(16)$ &15& 9.91 & 9.93 & -56.51 & -55.22 & 4.19 & 4.46\\
    \end{tabular}
  \end{center}
\end{table}

\section{Discussion}

Our Thiele simulations and analysis, along with other studies \cite{GobelElectricalwritingdeleting2019}, suggest that the degree $Q=0$, $S(1)$, skyrmionium bag would move in the direction of current flow. This is the case for the CPP geometry but in the CIP geometry the $S(1)$ significantly deviates from the expect $+ \be_x$ direction. The $S(1)$ bag is clearly symmetric (see Fig. (\ref{fig1})) and has off diagonal $\bD$ tensor components 5000 times smaller then the diagonal components indicating that asymmetry is not the driver of this deviation. This was unexpected and requires further research beyond this study.

It is clear, from the theory and simulations, that for both CIP and CPP geometries $S(0)$ to $S(4)$ skyrmion bags could be mechanically sorted, for these material parameters. Our Thiele analysis provides a sufficient benchmark for estimating differences in deflection between various topological degrees and a strong estimate of at what topological degree binning becomes impractical. 

In conclusion, for lower degree skyrmion bags, S(0) to S(4), in both the CIP and CPP geometries, we see significant differences in deflection angle (see Table (\ref{tab:deflections}). These textures are ideal candidates for logic gates, neuromorphic computing, data storage and processing applications. The CPP geometry with it's lower power requirements for equivalent velocities is the better choice in this analysis.

\section*{Acknowledgements}
This work was supported in part by the UK Engineering and Physical Sciences Research Council (EPSRC) grant EP/M506473/1. DF acknowledges funding by the Research Councils UK Energy Programme (Grant No. EP/T012250/1). The Titan V GPU used for parts of this research was donated by the NVIDIA Corporation.

\bibliography{phd}

\end{document}